\documentclass[12pt]{iopart}  \usepackage{graphicx}
\usepackage{subfigure,cite} \usepackage{amsfonts,amssymb}

\newcommand{\text}[1]{\mbox{\scriptsize{#1}}}

\begin{document}

\title[Thermal Fluctuations of A Translocating RNA]{Amplitude and
  Frequency Spectrum of Thermal Fluctuations of A Translocating RNA
  Molecule}

\author{Henk Vocks$^{\dagger}$, Debabrata Panja$^*$ and Gerard
  T. Barkema$^{\dagger,\ddagger}$}

\address{$\dagger$ Institute for Theoretical Physics, Universiteit
Utrecht, Leuvenlaan 4,\\  3584 CE Utrecht, The Netherlands

$^*$Institute for Theoretical Physics, Universiteit van Amsterdam,
Valckenierstraat 65,\\ 1018 XE Amsterdam, The Netherlands

$^{\ddagger}$Instituut-Lorentz, Universiteit Leiden, Niels Bohrweg 2,
2333 CA Leiden,\\ The Netherlands}

\begin{abstract}
Using a combination of theory and computer simulations, we study the
translocation of an RNA molecule, pulled through a solid-state
nanopore by an optical tweezer, as a method to determine its secondary
structure. The resolution with which the elements of the secondary
structure can be determined is limited by thermal fluctuations.  We
present a detailed study of these thermal fluctuations, including the
frequency spectrum, and show that these rule out single-nucleotide
resolution under the experimental conditions which we simulated.  Two
possible ways to improve this resolution are strong stretching of the
RNA with a back-pulling voltage across the membrane, and stiffening of
the translocated part of the RNA by biochemical means.
\end{abstract}


\maketitle

\section{Introduction}
New developments in design and fabrication of nano\-meter-sized pores
and etching methods, in recent times, have put translocation at the
forefront of single-molecule experiments
\cite{expts1,expts2,expts3,expts4,expts5,expts6,expts7,expts8,expts9},
with the hope that translocation may lead to cheaper and faster
technology for the analysis of biomolecules. The underlying principle
is that of a Coulter counter \cite{coulter}: molecules suspended in an
electrolyte solution pass through a narrow pore in a membrane. The
electrical impedance of the pore increases with the entrance of a
molecule as it displaces its own volume of solution. By applying a
voltage across the pore, the passing molecules are detected as current
dips. For nanometer-sized pores (slightly larger than the molecule's
cross-section) the magnitude and the duration of these dips have
proved to be rather effective in determining the size and length of
these molecules \cite{rob}.
\begin{figure}[h]
\begin{center}
\includegraphics[width=0.6\linewidth,angle=0]{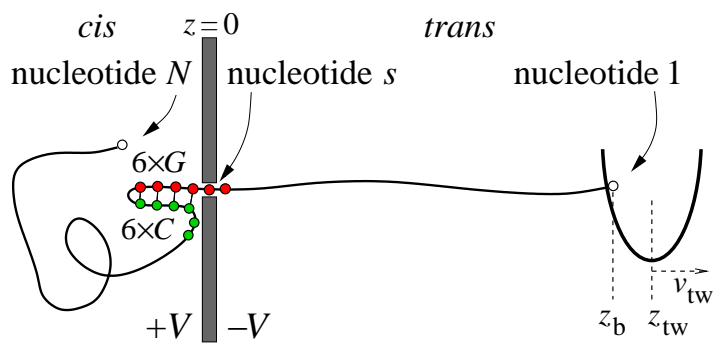}
\end{center}
\caption{The experiment in schematics, illustrated by only 6
$CG$-bonds for clarity. An RNA molecule composed of $N$ nucleotides is
pulled through a solid-state nanopore in a membrane (placed at $z=0$)
towards the right ($z>0$) using an optical tweezer, represented by a
parabolic potential, at a constant speed $v_{\text{tw}}$. The bottom
of the potential is located at $z_{\text{tw}}$; and the latex bead is
located at $z_{\text{b}}$. The number of monomers located on the {\it
trans}-side of the membrane is called $s$. The monomers are numbered,
starting from the end which is attached to the latex bead;
consequently, the nucleotide located in the pore is labelled $s$. A
potential difference $2V$ is also applied across the
membrane.\label{geometry}}
\end{figure}

With a membrane placed at $z=0$, a translocation experiment for
determining the secondary structure of an RNA molecule
\cite{pullingexp1,pullingexp2,pullingexp3,pullingth1,pullingth2,pullingth3,pullingth4}
composed of $N$ nucleotides proceeds as follows (see Fig.
\ref{geometry}). One end of the folded RNA, which is almost completely
located on the left ({\it cis\/}) side $(z<0)$ of the membrane, is
pulled through a solid-state nanopore to the right ({\it trans\/})
side  $(z>0)$, and a latex bead is attached to nucleotide 1. An
optical  tweezer captures the bead, and pulls the RNA through the pore
at a  constant speed $v_{\text{tw}}$. The monomer number at any given
time  located inside the pore is denoted by $s$. We denote the
voltages on the  {\it cis\/} and {\it trans\/} side by $+V$ and $-V$
respectively; a potential difference  $2V$ is thus applied across the
membrane, to increase the tension in the  translocated part of the RNA
so that no secondary structure can form  between the tweezer and the
pore. During this process, the force exerted  by the tweezer on the
RNA is monitored. Since the pore is narrow, for  translocation to
proceed, the bonds between the basepairs forming the  secondary
structures must be broken at the pore. The breaking of the
basepair-bonds is detected as increased force on the tweezer. The
force  on the tweezer as a function of time can then be translated
into the  binding energies of the basepairs as a function of distance
along the  RNA, yielding a wealth of information on the secondary
structure of the  RNA.

Note that actual RNA chains may be pulled through the pore either from
the $3'$ or from the $5'$ end, and the respective force extension
curves may pick up both the initial and final location of the stems.
From both the force-extension curves, with the help of a probabilistic
sequence alignment algorithm \cite{pullingth2}, one can subsequently
reconstruct the base-pairing pattern, the success of which clearly
depends on the accuracy with which the arrests in the force-extension
curves --- as experienced by the tweezer bead --- can be tied to the
actual translocation coordinates along the backbone of the RNA.

Given that the distances are of nanometer-scale in this experiment,
thermal fluctuations of the polymer are expected to blur the arrests
of the force extension curves; i.e., blur the coherence between the
force exerted by the tweezer and the nucleotide number located in the
pore. Since in this experiment one cannot track the events at the
pore, the unpredictability of the amount of low-frequency noise in
solid state nanopores seem to be the main barrier to progress in this
field \cite{noise1,noise2}. The central question addressed in this
paper, therefore, is the level of resolution (in units of a
nucleotide) that can be achieved by this experiment. All throughout
this paper, we define resolution as the accuracy with which the
location of secondary structure can be determined along the backbone
of the RNA --- which is strongly affected by the coherence between the
force exerted by the tweezer and the nucleotide number located in the
pore. We address this question by studying the amplitude and frequency
spectra of the fluctuations at the pore with a combination of theory
and computer simulations. The amplitude spectrum determines the
resolution limit that can be achieved by ensemble averaging, while
frequency spectrum determines the resolution limit that can be
achieved by time averaging. Note that the highest resolution that this
setup can achieve depends on which of these two limits is higher; this
prompts the study of both the amplitude and the frequency spectra of
the fluctuations at the pore. Our study rules out single-nucleotide
resolution under the experimental conditions which we simulated. Two
possible ways to improve this resolution are strong stretching of the
RNA with a back-pulling voltage across the membrane, and stiffening of
the translocated part of the RNA by biochemical means.

A related problem was considered by Thompson and Siggia
\cite{pullingth1}, who studied whether a measurable signal can be
obtained by pulling apart a DNA or RNA molecule by an atomic-force
microscope. They formulated their theoretical analysis using an
(equilibrium) partition sum that involved the interaction energy
between the unzipped strands. In the case of pulling apart a molecule
by translocation, the unzipped strands of the molecule are separated
by an impenetrable membrane, so the unzipped strands cannot interact
directly unless one of the strands translocates through the pore;
i.e., the process of unzipping {\it cannot\/} be decoupled from the
dynamics of translocation. Consequently, the method of Thompson and
Siggia cannot be easily imported to study our setup. We also note that
recently a number of researchers, e.g., Sauer-Budge {\it et al.}
\cite{sauer} and Bockelmann {\it et al.} \cite{bock} have studied the
case of pulling apart a molecule by translocation: however, their
formulations do not take into account any dynamics of translocation,
and therefore they only provide a simplified analysis of the
problem. Given that the (anomalous) dynamics of translocation involves
long memory effects \cite{anomlong,anom,planar,pulled,field}, we
follow a different method in this paper; this allows us to study the
full dynamic problem (i.e., including the frequency spectrum of the
thermal fluctuations).

Our work is related to the study of Bundschuh \etal\/
\cite{pullingth3} of translocation of RNA or DNA through a nanopore,
describing slow and fast regimes of translocation: for the former, the
{\it cis\/} side of the RNA molecule essentially remains equilibrated
at almost all times, while for the latter, the base-pairing pattern on
the {\it cis\/} side is essentially frozen during unzipping. Our
analysis describes a maximum pulling velocity of the optical tweezer
that allows the {\it trans\/} side of the molecule enough time to
always remain in its steady state, thereby providing the quantitative
distinguishing characteristics between the two regimes described by
Bundschuh \etal.

The structure of this paper is as follows. In Sec. \ref{sec2}, we
describe our computer model. In Secs. \ref{sec3} and \ref{sec4}, we
analyse the problem without and with thermal fluctuations
respectively. In Sec. \ref{sec5} we conclude the paper with a
discussion on the results.

\section{Computer model\label{sec2}}

We model the RNA with $N$ nucleotides as a lattice polymer with $N$
monomers on a face-centred-cubic lattice. Multiple occupation of the
same site is forbidden, i.e., the polymer is described by a
self-avoiding-walk. For practical purposes, this restriction is lifted
for consecutive nucleotides along the chain. The dynamics of the
polymer consists of single-nucleotide hops to nearest-neighbour sites,
attempted at random with rate unity, and accepted with Metropolis
probabilities. This model \cite{md1,md2,md3,md4} describes both
reptation and Rouse moves, but does not include explicit
hydrodynamics. We have used this model successfully to simulate
polymer translocation under various circumstances
\cite{anomlong,anom,planar,pulled,field}. Since our model is a variant
of a freely-jointed-chain, we expect it to reproduce poly(U) RNA
behaviour reasonably well \cite{fjc}.

In translocation experiments with biological nanopores, e.g.,
alpha-haemolysin, the polymer might show sequence-specific binding and
unbinding to the pore wall \cite{bez,expts6,expts7}. Such interactions
between the polymer and the membrane are not expected to play a role
in experiments of translocation through a solid-state nanopore, as
used in the experiments of Refs. \cite{expts8,expts9}. It has also
been suggested that the translocation of single-stranded DNA through
alpha-haemolysin nanopore is direction specific \cite{mathe} ($3'$ to
$5'$ as opposed to $5'$ to $3'$); in the same paper, by computer
simulations, the authors demonstrated that such direction-specificity
should not be present when the pore diameter is $\gtrsim1.5$ nm. Given
that the typical diameters for solid-state nanopores are $\gtrsim2$ nm
\cite{poresize}, in our simulations, we neglect interactions between
the polymer and the membrane, other than excluded-volume interactions
(i.e., the polymer cannot cross the membrane other than through the
pore).

Since we want to study how secondary structure influences the
translocation process, we add the ability for parts of the polymer to
form two hairpins. The real RNA sequence which comes closest to our
approach is a poly(U) RNA with a sequence composition $U_{30} \left(
U_{60} G_{32} U_6 C_{32} \right)_2 U_{60}$, wherein {\it each
nucleotide corresponds to a monomer}: two $C$ and $G$-nucleotides on
neighbouring lattice sites can form a bond with an affinity $E_{CG} =
2.3\,k_{\mathrm{B}}T$, but we do not allow $GU$ pairing. The latter is
a simplification from how a real RNA molecule with the above sequence
would behave, but with this simplification we {\it a priori} know what
to expect for the secondary structure of this polymer --- namely two
hairpins, each with $32$ $CG$-bonds \cite{notehairpin} --- on which we
study the effect of thermal noise that limit the achievable basepair
resolution of the secondary structures.

Our model of the optical tweezer is that the latex bead, i.e.,
nucleotide 1 feels a spherically symmetric harmonic trap with spring
constant $k_{\mathrm{tw}}$, centred around the location of the optical
tweezer at a distance $z_{\mathrm{tw}}$ from the membrane.

It is clear that our model does not capture the full details of a real
laboratory experiment: indeed a more detailed model could include
explicit hydrodynamics, detailed RNA interactions such as GU pairing,
and a more elaborate description of the charge distributions on the
RNA. Leaving out explicit hydrodynamics does alter polymeric motion
(from Zimm to Rouse dynamics) and therefore is likely to affect the
polymer's memory effects. Although at this moment we do not precisely
know how the memory effects that are relevant for the present problem
--- discussed in Sec. \ref{sec4} --- will alter when explicit
hydrodynamics are incorporated, the low-frequency domination of the
memory effects will not disappear.

To correspond to experimental parameters we use a lattice spacing of
$\lambda=0.5\,\mbox{nm}$, comparable to the persistence length (as
well as the typical inter-nucleotide distance) for poly(U)
\cite{seol}. The resulting forces measured at the tweezer are
$60\,\mathrm{pN}$ or less, similar to experimental values
\cite{seol}. Equating the diffusion coefficient $2$-$5\times 10^{-6}$
cm$^2$/s for $U_6$ \cite{difRNA} to that of a polymer of length $N=6$
in our model, one unit of time in our simulations corresponds roughly
to $100\,\mbox{ps}$. The tweezer velocity is one lattice spacing
$\lambda$ per $300,000$ units of time, or $\sim 20\,\mu$m/s,
comparable to typical experimental velocities. The time scale
$\lambda/v_{\text{tw}}$ is larger than that (or of the order) of the
longest time-scale of the translocated part of the RNA, implying that
{\it the translocated part of the RNA can be treated as properly
thermalised at all times.}  In our simulations, the value of $\Delta U
= 2qV$ ranges from 0.4 to 2.75 $k_{\mathrm B}T$.  Given that at room
temperature $k_{\mathrm B}T=25$ meV and assuming that each nucleotide
carries an effective charge around $q=0.5$ times the electron charge
(due to Manning condensation, which limits the charge to one electron
charge per Bjerrum length\footnote{Due to Manning condensation, the
effective charge per unit length is limited to approximately one
electron charge per Bjerrum length. In water, the Bjerrum length is
about 7 \AA. Since the typical RNA base pair distance is
$\approx3.4$\AA, the effective charge is approximately $0.5e$ per
nucleotide. In the pore, the dielectric constant is significantly
lower than that of water, consequently the Bjerrum length (which is
inversely proportional to the dielectric constant) is much larger, and
hence the effective charge is much lower (a tenth of an electron
charge per nucleotide or even less \cite{zhang,sauer}). For our work,
a key quantity is the stretching force, determined by the energy
difference across the pore, set by the effective charge in
solution. Thus, for our purpose, the relevant effective charge is 0.5e
per nucleotide. The main consequence of the much lower effective
charge inside the pore is that RNA is less eager to enter the pore,
but that has no consequences for this work.}) \cite{poon}, our
simulations correspond to an experimentally applied voltage
differences ranging from $10\,\mbox{mV}<V<70\,\mbox{mV}$.

\begin{figure}[h]
\begin{center}
\includegraphics[width=0.5\linewidth,angle=270]{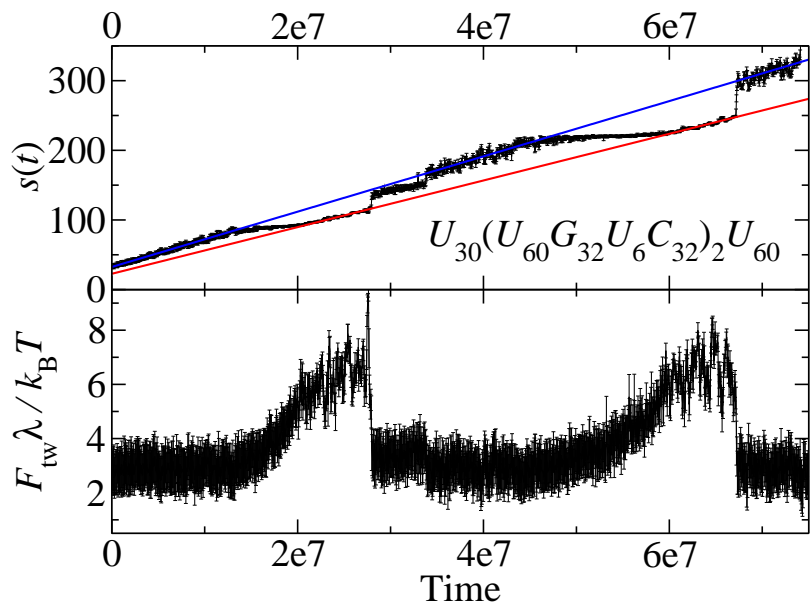}
\end{center}
\caption{Upper panel: nucleotide in the pore $s$ as a function of time
for a poly(U) RNA of composition $U_{30} \left( U_{60} G_{32} U_6
C_{32} \right)_2 U_{60}$, pulled with constant velocity
$v_{\mathrm{tw}}=\lambda$ per $300,000$ time steps (approximately 20
$\mu$m/s). The binding energy of each bond is set to $E_{CG}=
2.3\,k_{\mathrm{B}}T$, with $2qV = 1.5\,k_{\mathrm{B}}T$. Every data
point is an average of $1500$ consecutive measurements each $100$ time
steps apart, with the standard deviation represented by the error
bars. The two straight lines are guides for the eye. Lower panel: the
corresponding chain tension measured by means of the optical tweezer,
with $k_{\mathrm{tw}} = 1\,k_{\mathrm{B}}T/\lambda$.\label{unzipping}}
\end{figure}

A typical simulation output is presented in Fig. \ref{unzipping}. It
consists of the force $F_{\mathrm{tw}}(t)$ exerted by the optical
tweezer as a function of time: {\it this information is readily
accessible in real experiments}. In our simulations, we also monitor
the number $s$ of monomer located in the pore as a function of time:
{\it this information is typically not accessible in real
experiments}.  In the simulation of figure \ref{unzipping}, the force
exerted by the tweezer hovers around a fixed strength
$3.4\,k_{\mathrm{B}}T/\lambda$ (approximately 30 pN), except between
$t=1.4\times10^7$ to $3.4\times10^7$ resp. $t=4.6\times10^7$ to
$6.7\times10^7$, when the first and second hairpins are pulled through
the pore. Indeed, the top panel of Fig. \ref{unzipping} shows that at
the onset of these intervals, $s(t)$ is almost constant, around $90$
and $220$, the starting locations of the hairpins.

\section{Translocation without thermal fluctuations\label{sec3}}

First we discuss what sort of information on the secondary structure
can {\it ideally\/} --- i.e., in the {\it absence\/} of thermal
fluctuations --- be obtained from $F_{\mathrm{tw}}(t)$. We do this
under the assumptions that the force extension curve of the RNA
without any secondary structure is sequence-independent, and that the
tweezer velocity is low enough for the force exerted by the tweezer to
maintain a uniform chain tension $\phi$ all along the translocated
part of the RNA. Then, $\phi$ is uniquely determined by the relative
extension $x=z_{\mathrm{b}}/s$, and is balanced by
$F_{\mathrm{tw}}(t)$, i.e.,
\begin{equation}
\phi=\mathcal{F}(x)=F_{\mathrm{tw}}=k_{\text{tw}}(z_{\text{tw}}-z_{\text
b}),
\label{rel1}
\end{equation} where $k_{\text{tw}}$ is the stiffness of the optical
tweezer.  Additionally, the equality of the rate of work done by the
tweezer and the gain in free energy by the translocating nucleotides
at the pore yields
\begin{equation} F_{\mathrm{tw}}\,dz_{\mathrm{b}}=\left(\Delta U -
T\Delta S\right)ds,
\label{rel2}
\end{equation} where $\Delta S$ is the entropic cost per nucleotide
translocation due to the imbalance of the (entropic) chain tension
across the pore, and $\Delta U$ is the energetic cost per nucleotide
translocation. If translocation of the nucleotides does not involve
breaking of $CG$-bonds at the pore, then $\Delta U=\Delta
U_c\equiv2qV$, otherwise $\Delta U=\Delta U_b\equiv2qV+E_{CG}$. Thus,
given $z_{\text{b}}$ and $\left(\Delta U-T\Delta S\right)$,
Eqs. (\ref{rel1}) and (\ref{rel2}) determine both the tweezer force
and the relative extension.

During the translocation of the first 90 nucleotides of $U_{30}
\left(U_{60} G_{32} U_6 C_{32} \right)_2 U_{60}$, no secondary
structure is broken at the pore --- consequently, $\Delta U=\Delta
U_c$ --- and the tweezer force remains constant at
$F_{\mathrm{tw}}(t)=3.4\,k_{\mathrm{B}}T/\lambda$ (approximately 30
pN). The speed of translocation is then given solely by
Eq. (\ref{rel1}), with $\dot z_{\text{b}}=v_{\text{tw}}$, as
$\dot{s}=v_{\mathrm{tw}}/\mathcal{F}^{-1}[F_{\mathrm{tw}}(t)]=1$
nucleotide per $252,000$ time steps, for which we have used the
numerically obtained force extension curve (inset, Fig. \ref{figds2});
this speed $\dot s$ is shown in Fig. \ref{unzipping} by the upper
(blue) line.

Following the arrival of the first hairpin at the pore, translocation
requires breaking of the $CG$-bonds, and consequently, $\Delta U$
increases by $E_{CG}=2.3\,k_{\mathrm{B}}T$. Both the tweezer force and
the relative extension adjust to new values, determined by
Eqs. (\ref{rel1}) and (\ref{rel2}) for the new value for $\Delta U$.
The resulting tweezer force equals
$F_{\mathrm{tw}}(t)=7.5\,k_{\mathrm{B}}T/\lambda$ (approximately 70
pN); the correspondingly adjusted speed $\dot s$ is shown in
Fig. \ref{unzipping} by the lower (red) line.

After the translocation of the first $32$ $G$-nucleotides, $\Delta U$
returns to its base value $\Delta U_c$. The tweezer force and the
relative extension, too, fall back to their pre-hairpin values. This
is seen in Fig. \ref{unzipping} by $s(t)$ leaving the lower red line
{\it sharply\/} to re-coincide with the upper blue line; i.e., quite a
few nucleotides at the end of the hairpin translocate nearly
immediately. This chain of events repeats itself during translocation
of the second hairpin: first the tweezer force increases gradually to
its higher value and the translocated distance approaches the lower
red line, then the tweezer force decreases steeply to its lower value
and the translocated distance jumps to the upper blue line.

In conclusion, most features of Fig. \ref{unzipping} are qualitatively
well-understood.  The above framework can be easily extended to a
wider set of bond strengths and more elaborate secondary
structures. Without thermal fluctuations, the setup is perfectly
suited to determine the secondary structure up to the nucleotide
resolution, under the restriction that the consecutive bonds along the
backbone of the RNA are of increasing strength; if strong bonds are
followed by weaker bonds that are not strong enough to halt the
translocation process, the breaking of the weaker bonds will not be
accompanied by an increase of $F_{\text{tw}}$ and the experiment will
reveal little information about these weaker bonds.

\section{Translocation with thermal fluctuations\label{sec4}}

In reality, thermal fluctuations are omnipresent in this nano-scale
experiment, and as argued earlier, for solid-state nanopores they are
the dominant source of noise at the pore. In fact, it is precisely the
(thermal) fluctuations in $s(t)$ that serve to blur the coherence
between $F_{\text{tw}}(t)$ and $s(t)$, and thereby limit the
resolution that can be achieved by this experiment. We will now study
both the amplitude and the frequency spectrum of these thermal
fluctuations in $s(t)$.
\begin{figure}[h]
\begin{center}
\includegraphics[width=0.4\linewidth,angle=270]{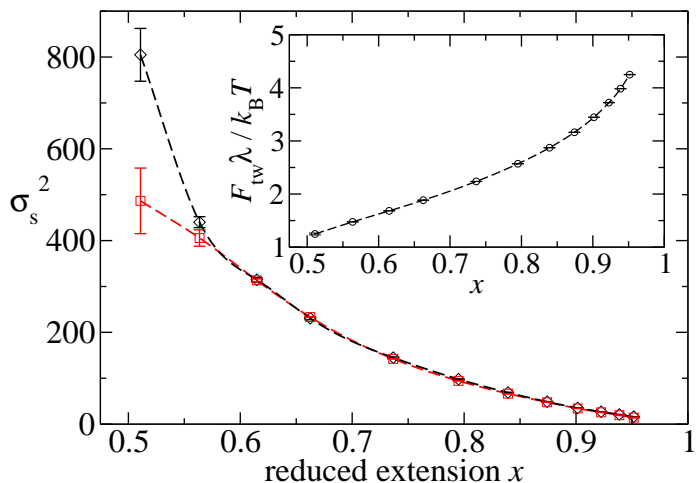}
\end{center}
\caption{$\sigma_s^2$, the mean square displacement of $s$ vs. reduced
extension $x=z_b/s$, at constant $z_{\text{tw}}=300\,\lambda$. The
chain tension $\phi$ is slowly increased by changing $\Delta U=2qV$
from $0.40$ to $2.75\,k_{\mathrm{B}}T$. Each data point required $80$
independent polymers and simulation times of $20$ million time steps,
with a measurement every $100$ time steps. Data points from direct
simulations and Eq. (\ref{ds2}) are represented as black diamonds and
red squares. The dashed lines are cubic splines. The error bars
represent statistical errors only. Inset: Rescaled force-extension
curve for our model.\label{figds2}}
\end{figure}

The amplitude $\sigma_s(t,\Delta U)$ of the fluctuations in $s(t)$ is
that of an entropic spring at fixed extension $z_{\text b}$ with one
end tethered at the tweezer, while the number of nucleotides in the
spring are allowed to fluctuate through the pore. Now consider a {\it
different\/} problem --- an entropic spring with an average, but
fluctuating extension $z_{\text b}$. From the equipartition theorem,
these fluctuations are given by $\langle\delta z_{\text
b}^2\rangle=k_{\mathrm{B}}T/c_{z_{\text b}}$ [with spring constant
$c_{z_{\text b}}=(\partial\mathcal{F}/\partial z_{\text
b})_s=\mathcal{F}'(x)/s$, in which $x=z_{\text b}/s$ is the relative
extension]. For the present problem, such fluctuations in $z_{\text
b}$ can be thought of to be mediated by the fluctuations in $s$,
yielding \cite{eqremark}
\begin{equation} \sigma_s^2(t,\Delta U)=\langle\delta
z_b^2\rangle\left[\frac{\partial s}{\partial z_{\text
b}}\right]_x^2=s(t)\!\left[x^2 \mathcal{F}'(x)\right]^{-1}
k_\mathrm{B}T.
\label{ds2}
\end{equation} 

In Fig. \ref{figds2}, Eq. (\ref{ds2}) is compared to
the simulation results for several values of $\Delta U$, with constant
$z_{\mathrm{tw}} = 300\,\lambda$. Note that $\Delta U$ only serves to
set the value of $\phi$. Of practical importance is the observation
that according to Eq. (\ref{ds2}) the amplitude of the thermal
fluctuations decreases with increasing tension. Thus, an increase of
the applied voltage difference will reduce the thermal fluctuations in
$s$ thereby increasing the resolution of the secondary structure
determination. The same effect can also be achieved by increasing the
chain stiffness, thereby increasing $\mathcal{F}'(x)$. In practice,
this could for instance be realised through chemical means. For
instance, it is known that the salinity affects the chain stiffness
\cite{seol}.  Also, certain proteins (such as RecA for ssDNA) can be
added on the trans-side of the membrane, to increase the chain
stiffness dramatically, while at the same time reducing secondary
structure formation on the side of the tweezer.
\begin{figure}[h]
\begin{center}
\includegraphics[width=0.45\linewidth,angle=270]{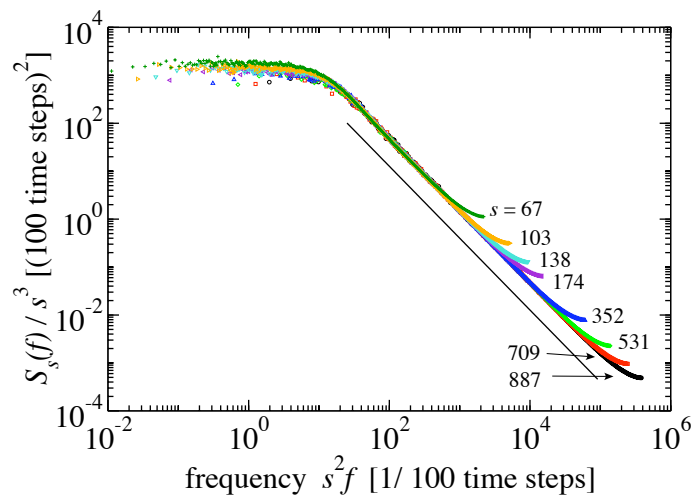}
\end{center}
\caption{(Rescaled) power spectrum of $s\left(t\right)$,
$S_s\left(f\right)$, versus (rescaled) frequency for $\Delta U = 1.5$
$k_{\mathrm{B}}T$, and $z_{\mathrm{tw}}/\lambda$=$60$, $90$, $120$,
$150$, $300$, $450$, $600$, $750$. Each curve is composed of
statistics from $80$ polymers for 40 million time steps, the value of
$s$ corresponding to each curve are shown in the Figure. The solid
line $\sim f^{-3/2}$ is added as a guide to the
eye. The unit of $s^2f$ along the horizontal axis, 1/100 time steps, is approximately
equal to 100 MHz.\label{fluctuations}}
\end{figure}

For the frequency spectrum of $s(t)$, we return to
Ref. \cite{pulled}. Therein we showed that $\dot s(t)$ and the chain
tension imbalance across the pore are related via a time-dependent
memory kernel $a(t)$. This result, adapted to the notations in this
paper, is given by
\begin{equation} \dot{s}(t)=\int_{-\infty}^t
dt'\,a(t-t')\,[\phi(t')-\phi_{z<0}(t')],
\label{e4}
\end{equation} 
where $\phi_{z<0}$ is the chain tension of the RNA at
the $z<0$ side of the pore, and $a(t) \sim t^{-3/2}
\exp[-t/\tau_{\mathcal{F}}]$; thus, the memory kernel $a(t)$ is
described by a power-law decay with exponent $-3/2$, up to some
cut-off time $\tau_{\mathcal{F}}$; this cut-off time is found to
increase quadratically with the length of the translocated RNA:
$\tau_{\mathcal{F}} \sim s^2$ \cite{pulled}. The immediate consequence
of Eq. (\ref{e4}) is that
\begin{equation} 
\langle \dot{s}(t)\dot{s}(t')\rangle \sim
|t-t'|^{-3/2}\exp\left[-|t-t'|/\tau_{\mathcal{F}}\right],
\label{alpha}
\end{equation} 
following the fluctuation-dissipation theorem \cite{noteeqvel}, i.e.,
\begin{eqnarray}  
\hspace{-4mm}\langle s(t)s(t') \rangle =
\int_0^tdt_1\int_0^{t'}dt_2\langle \dot{s}(t_1)\dot{s}(t_2)\rangle
\nonumber\\&&\hspace{-6cm}=\left\{\begin{array}{ll}
4\left[|t-t'|^{1/2}-|t|^{1/2}-|t'|^{1/2}\right] & \quad|t-t'| \lesssim
\tau_{\mathcal{F}}\\ C  & \quad|t-t'|\gg \tau_{\mathcal{F}}
\end{array}\right.,
\label{autocorv}
\end{eqnarray} 
with some constant $C$. This autocorrelation function,
for $|t-t'| \lesssim \tau_{\mathcal{F}}$, has the same form as that of
a fractional Brownian motion (FBM) \cite{PRE62_6103andref,hurstref}
\begin{equation}  
\langle s(t)s(t')\rangle=
C_1\left(|t|^{2H}+|t'|^{2H}-|t-t'|^{2H}\right),
\label{e9}
\end{equation}  
with Hurst parameter $H=1/4$, and some constant $C_1$.
Since the r.h.s. of Eq.~(\ref{e9}) is not a function of $(t-t')$
alone, the spectral density of the FBM is not well-defined. However,
by applying generalised concepts such as the Wigner-Ville spectrum
\cite{PRE62_6103andref}, a limiting power spectrum can be obtained as
follows: $S_s(f)=\frac{C_2}{|f|^{\alpha}}$, with $\alpha=2H+1=3/2$,
with some constant $C_2$.  We therefore expect a power spectrum, with
$f_c\sim 1/\tau_{\mathcal{F}} \sim 1/s^2$,
\begin{equation}  S_{s}(f) =\left\{ \begin{array}{ll}
\frac{C_2}{|f|^{3/2}} & (|f| > f_c) \\ \frac{C_2}{f_c^{3/2}} & (|f| <
f_c)
\end{array}\right..
\label{powspec}
\end{equation}   

As shown in Fig. \ref{fluctuations}, after having
rescaled $S_{s}(f)$ and $f$ by $s^{-3}$ and $s^2$ respectively, the
collapse of the power spectra confirms Eq. (\ref{powspec}). The
spectrum is dominated by the low frequency fluctuations of $s(t)$,
which are not easily suppressed by time averaging.

The translocated part of the RNA can be considered to be in a steady
state as long as $\dot{s}(t)<\tau_{\mathcal{F}}^{-1}$, i.e., the time
between the passage of consecutive nucleotides through the pore does
not exceed the time for the longest memory in the translocated part of
the RNA. This time for the longest memory is set by the longest length
$s_{\mathrm{max}}$ of the translocated  part of the RNA. If the
tweezer is pulled with a velocity exceeding
\begin{equation} v_{\mathrm{tw},\mathrm{max}}(t,\Delta U)=x\cdot
f_c\sim\frac{C_2^{\,2}x^5\mathcal{F}'(x)^2}{s_{\mathrm{max}}^2(k_{\mathrm{B}}T)^2},
\end{equation} where Parseval's theorem has been used to equate
Eq. (\ref{powspec}) to Eq. (\ref{ds2}), the {\it trans\/} side of the
molecule is not allowed enough time to reach the steady state. This
results in forces at the tweezer that far exceed the forces required
to pull monomers through the pore at steady state (of the molecule on
the {\it trans\/} side). Fig. 4 shows that if a potential difference
of $\Delta U = 1.5k_{\mathrm{B}}T$ is applied, the power rescaled
spectrum crosses over from a frequency-independent regime to a
power-law regime if $s^2f=0.1$ to 0.15; with a pulling velocity
$v_{\mathrm{tw}}=\lambda/300,000$ (roughly $\sim 20\,\mu$m/s) the
typical frequency is $f\sim 300,000^{-1}\approx$ 0.3 MHz; the rescaled quantity
$s^2f$ then reaches the crossover once the translocated part of the
chain has a length $s_{\mathrm{max}}\approx 200$ nucleotides. Beyond this length,
the translocated part of the chain will be pulled out of its
equilibrium shape, and the force felt by the tweezer is no longer
determined by what happens near the pore, but rather by what happens
in the translocated part of the chain, with the consequence that
limited information about possible secondary structure will be
obtained.  In targeted simulations we have indeed observed the
behaviour that {\it without\/} any base-pairing, for tweezer
velocities higher than $\lambda/300,000$ ($\approx$20 $\mu$m/s), the force at the tweezer is
an increasing function of time significantly once
$s_{\mathrm{max}}\approx 200$ has been reached.

\section{Discussion and conclusions\label{sec5}}

Having demonstrated the importance of thermal fluctuations in $s(t)$,
let us now revisit the translocation of $U_{30} \left( U_{60} G_{32}
U_6 C_{32} \right)_2 U_{60}$, cf. Fig. \ref{unzipping}. Except between
times $1.4$ and $3.4\times10^7$ (resp. $4.6$ and $6.7\times10^7$),
$F_{\mathrm{tw}}(t)$ is roughly $3.4\,k_{\mathrm{B}}T/\lambda$ (approximately 30 pN), with
fluctuations $\langle\Delta F_{\text{tw}}^2(t)\rangle\approx
k_{\mathrm{tw}} k_{\mathrm{B}}T$, while Eq. (\ref{ds2}) explains the
growth of $\sigma_s^2(t)$ linearly with $s(t)$ (hence linearly with
$t$). It is important to stress that while the fluctuations in
$F_{\text{tw}}(t)$ may be removed to a reasonable degree by time
averaging (e.g., by reducing the velocity of the tweezer), the
fluctuations in $s(t)$ fundamentally reduce the accuracy with which
RNA secondary structure can be observed. As soon as the hairpin is
within a distance of order $\sigma_s(t,\Delta U)$ from the pore, i.e.,
$\langle s(t)\rangle +\sigma_s(t,\Delta U_c) =
90\,(\mbox{resp.}\,\,220)$, the tweezer force increases already, to
reach a plateau once $\langle s(t)\rangle-\sigma_s(t,\Delta U_b)$ $=$
$90\,(\mbox{resp.}\,\,220)$. Translocation at the pore then proceeds
with a constant rate
$v_{\mathrm{tw}}/\mathcal{F}^{-1}(F_{\mathrm{tw}})$, until $\langle
s(t)\rangle + \sigma_s(t,\Delta U_b)=122\,(\mbox{resp.}\,\,252)$ (in
which we assume that the effective affinity $E_{CG}$ is not altered
much as the unzipping of the hairpin proceeds), at which time the
chain tension is released by rapid translocation of the remaining
nucleotides.  These considerations clearly establish that only
secondary structure elements with at least two times
$\sigma_s(t,\Delta U_b)$ basepairs can be detected accurately,
assuming that the affinities of consecutive basepairs are more or less
alike.  With the choice of parameters in the simulations leading to
Fig. \ref{unzipping} which are comparable to typical experimental
values, this means that the resolution is limited to $\approx 8$
nucleotides.  Information on smaller lengths is washed out by the
thermal fluctuations, making it very hard to retrieve.
Moreover, the basepairs of the preceding stem of the
secondary structure must be characterised either by weaker affinities,
or by strong heterogeneity.

Although the fluctuations in $s(t)$ are not easily suppressed,
Eq. (\ref{ds2}) does leave open avenues for higher accuracy, either by
increasing the relative extension $x=z_b/s$, or by increasing the
stiffness $\mathcal{F}'(x)$ of the polymer on the trans side of the
membrane. The former can be achieved by applying a stronger potential
difference $2V$ (the strength of which is of course limited by
experimental considerations). On the other hand, the polymer's
stiffness can be actively enhanced, e.g., by altering the salinity of
the solution \cite{seol}; alternatively, if one is interested in doing
single-molecule experiments with ssDNA, the addition of RecA proteins
to the solution only on the trans of the membrane may be of help as
well (this will increase the stiffness of the polymer on the trans
side of the membrane, but since these proteins cannot pass through the
pore, the secondary structure on the cis side will be left
unaffected).

If biological pores such as alpha-haemolysin are used for this
experiment instead of solid-state nanopores, the translocating polymer
will show sequence-dependent binding/unbinding to the pore
wall. Although we cannot oversee all consequences of such
interactions, we do not expect these to affect the coherence between
the force felt at the tweezer and the translocated length, nor the
frequency spectrum of the fluctuations in the translocated part of the
chain. Hence, our conclusions should also apply to biological pores.

\section*{Acknowledgement} We acknowledge useful discussions with our
experimental colleagues Cees Dekker and Nynke Dekker at Delft
University of Technology, The Netherlands. D. P. gratefully
acknowledges ample computer time at the Dutch national supercomputer
facility SARA.


\begin{thebibliography}{10}
\bibitem{notehairpin} Note here that the sequence $U_{30} \left(U_{60}
G_{32} U_6 C_{32} \right)_2 U_{60}$ has a prominent alternative
structure (besides two hairpins): one long hairpin with a loop of size
$60$. Such a loop of size $60$ is often observed within our polymer
model. However, in our polymer model secondary structures tend to be
highly compactified, which makes it difficult to provide a
quantitative comparison with real RNA chains.
\bibitem{eqremark} The direct simulation results and Eq. (\ref{ds2})
in Fig. \ref{figds2} deviate from each other for $x\lesssim0.6$, since
in this case the assumption  $\sigma_s \ll s$ does not hold any
more. Also, tethering of nucleotide 1 requires $k_{\mathrm{tw}} \gg
c_{z_b}$; this assumption is clearly valid in our simulations, but
need not be valid in real experiments.
\bibitem{noteeqvel} Note here that Ref. \cite{pulled} is about pulling
the tweezer with a constant force, while in this paper the tweezer is
pulled with a constant velocity.  This does not affect the memory
kernel and the consequences of the RNA's memory.
\end{thebibliography}
\end{document}